\documentclass[amssymb,a4style,superscriptaddress]{revtex4}
\usepackage{amsmath}
\usepackage{amssymb}
\usepackage{amsmath}
\usepackage{amsfonts,amssymb,amstext}

\font\sqi=cmssq8
\def\DR{\rm I\kern-1.45pt\rm R}
\def\DC{\kern2pt {\hbox{\sqi I}}\kern-4.2pt\rm C}
\newcommand{\poiss}[2]{\{#1,#2\}}

\font\sqi=cmssq8
\def\DR{\rm I\kern-1.45pt\rm R}
\def\DC{\kern2pt {\hbox{\sqi I}}\kern-4.2pt\rm C}
\newcommand{\ba}{\begin{array}}
\newcommand{\ea}{\end{array}}
\newcommand{\be}{\begin{equation}}
\newcommand{\ee}{\end{equation}}
\newcommand{\bea}{\begin{eqnarray}}
\newcommand{\eea}{\end{eqnarray}}
\begin{document}
\begin{flushright}
ITP--UH--12/09
\end{flushright}
\title{Hidden symmetries of integrable conformal mechanical systems}
\author{Tigran Hakobyan}
\email{hakob@yerphi.am}
\affiliation{Yerevan State University, 1 Alex Manoogian St., 0025 Yerevan, Armenia}
\affiliation{Yerevan Physics Institute, 2 Alikhanyan Brothers St., 0036 Yerevan, Armenia}
\author{Sergey Krivonos}
\email{krivonos@theor.jinr.ru}
\affiliation{Bogoliubov  Laboratory of Theoretical Physics, JINR,
141980 Dubna, Russia}
\author{Olaf Lechtenfeld}
\email{lechtenf@itp.uni-hannover.de}
\affiliation{ Institut f\"ur Theoretische Physik, Leibniz Universit\"at Hannover,
30167 Hannover, Germany} 
\author{Armen Nersessian}
\email{arnerses@yerphi.am}
\affiliation{Yerevan State University, 1 Alex Manoogian St., 0025 Yerevan, Armenia}
\affiliation{ Artsakh State University, 5 Mkhitar Gosh St., Stepanakert, Armenia}
\begin{abstract}
\noindent
We split the generic conformal mechanical system into a ``radial'' and an
``angular'' part, where the latter is defined as the Hamiltonian system
on the orbit of the conformal group, with the Casimir function in the role
of the Hamiltonian. We reduce the analysis of the constants of motion
of the full system to the study of certain differential equations
on this orbit. For integrable mechanical systems, the conformal invariance
renders them superintegrable, yielding an additional series of conserved
quantities originally found by Wojciechowski in the rational Calogero model.
Finally, we show that, starting from any ${\cal N}{=}4$ supersymmetric 
``angular'' Hamiltonian system one may construct a new system with full 
${\cal N}{=}4$ superconformal $D(1,2;\alpha)$  symmetry.
\end{abstract}
\maketitle
\section{Introduction}

Conformal invariance plays an important role in many areas of
the quantum field theory and  condensed matter physics, especially in 
string theory, the theory of critical phenomena, low-dimensional
integrable models, spin and fermion lattice systems.
Therefore, one-dimensional (mechanical) systems with
conformal invariance are worth to be studied in detail
and still require further investigation.
Actually, the conformal group is not an exact
symmetry for the conformal mechanical system. It does not commute
with the Hamiltonian but, instead, is a symmetry of the action (a
symmetry in the field theoretical context). The Hamiltonian itself
forms an $so(1,2)$ algebra together with generators of the
dilatation  and conformal boost, with respect to canonical Poisson
brackets. It is interesting that, due to the conformal symmetry, 
the ``angular" part of the Hamiltonian of conformal mechanics is 
a constant of motion. However, its relation with other constants of motion
has not been investigated properly so far.

The (rational) Calogero model \cite{calogero69,calogero71,moser},
which is an integrable $N$-particle one-dimensional system with pairwise
inverse-square interaction (and its various generalizations related
with different Lie algebras and Coxeter groups~\cite{algebra})
is a famous example of a conformal mechanical system
(for the review, see~\cite{polychronakos}). 
Usually, Lax-pair and matrix-model approaches
are employed for the study of this system. These are common methods which 
are applied to other integrable models not related to the conformal group.
At the same time, many properties of the rational Calogero model are
due to its conformal invariance, and they are shared with 
other conformal mechanical models. For example, 
the ``decoupling transformation" in the Calogero model~\cite{hindu}
can be formulated purely in terms of conformal transformations~\cite{anton} 
(see also~\cite{lobach}). Note that the rational
$N$-particle Calogero model is a maximally superintegrable system, 
i.e.~it possesses $N{-}1$ additional
functionally independent integrals apart from the Liouville
integrals being in involution~\cite{supint}. Despite an impressive
list of references on this subject~\cite{supint2}, the
superintegrability of the Calogero model still seems to be mysterious.
Preliminary considerations have indicated a
direct connection between the additional constants of motion and the
``angular" part of the Calogero model~\cite{cuboct}. This observation
allows us to propose that conformal invariance provides any integrable 
conformal system with the superintegrability property. 
In the present paper we prove this statement.

The ${\cal N}{=}4$ superconformal extension of the $N$-particle Calogero system
based on the $su(1,1|2)$ superalgebra~\cite{sul} is intimately related with
solutions of the WDVV equation and therefore relevant for topological field
theory.
By looking at the ${\cal N}{=}4$ superconformal three-particle rational 
Calogero system~\cite{krivonos}, we find evidence 
that any ${\cal N}{=}4$ supersymmetric extension of the respective
angular Hamiltonian system could be lifted to some ${\cal N}{=}4$ 
superconformal mechanics.

In present paper we clarify these issues.
For this purpose we invariantly split the generic conformal mechanics into 
a ``radial"  and an ``angular" part, where the latter is a Hamiltonian system 
on the orbit of conformal group, with the Casimir function playing
the role of the Hamiltonian.
We investigate how the constants of motion of the conformal mechanics 
are encoded in this ``angular Hamiltonian system".
We reduce the analysis of the constants of motion to the study of
certain differential equations in the angular variables.
As one of the main results of this paper, we demonstrate that the
additional series of constants of motion of the Calogero model found 
by Wojciechowski can be constructed as Poisson brackets
of the Liouville constants of motion with the angular Hamiltonian of the
Calogero model.
Remarkably, these additional integrals exist for generic conformal mechanics. 
This means that any integrable conformal mechanics is also superintegrable. 
We give the explicit expression relating the constants of motion of the 
conformal mechanics with those of its angular part.
All these observations may easily be extended to the case of superconformal 
systems. As an example, in the last Section we explicitly
demonstrate that any ${\cal N}{=}4$ supersymmetric ``angular'' system can 
be lifted to some mechanics with full ${\cal N}{=}4$ superconformal 
$D(1,2;\alpha)$ symmetry.

\setcounter{equation}{0}
\section{The model}
Let us consider an arbitrary conformal mechanics defined by
the  Hamiltonian $H$ which forms, together with some generators of dilatations $D$ and conformal boost $K$
the one-dimensional conformal algebra $so(2,1)$
\be
\label{so12}
\{  H , D\}= 2 H,  \qquad  \{ K, D\}=-2 K, \qquad\{  H , K\}=D.
\ee
The Casimir element of
\eqref{so12} is given by the expression
\be
\label{casimir}
\mathcal{I}=2KH-\frac{1}{2}D^2
\ee
If we now  define the radial coordinate $r$ and its conjugated momentum $p_r$ as follows
\be
r\equiv {\sqrt{2K}},\qquad p_r\equiv\frac{D}{\sqrt{2K}}\; :\qquad\{p_r, r\}=1,
\label{rpr}\ee
then, taking into account that $\mathcal{I}$ is the Casimir of
of $so(1,2)$ with the explicit form  \eqref{casimir},
we will arrive at the system
\begin{equation}
H=\frac{{p}^2_r}{2}+\frac{\mathcal{I}}{r^2},\quad D=rp_r  ,
\quad
K=\frac{r^2}{2},
\qquad
\{p_r, r\}=1,\label{so2}
\end{equation}
where
\be
\{\mathcal{I}, H\}=0,\qquad \{p_r, \mathcal{I}\}=\{r, \mathcal{I}\}=0.
\ee
Thus, $\mathcal{I}$ is indeed {\it the constant of motion} of the Hamiltonian $H$.
Therefore, although conformal symmetry is not a symmetry of the Hamiltonian, it equipped
the system with the additional (to the Hamiltonian) constant of motion $\mathcal{I}$.

Summarizing we conclude that the generic mechanical system with 
dynamical symmetry given
by conformal algebra so(1,2) (\ref{so12}) is defined by the
Hamiltonian system
\be  \Omega=dp_r\wedge dr+\omega_0(u),\quad
H=\frac{{p}^2_r}{2}+\frac{\mathcal{I}(u)}{r^2} \label{generic}\ee
where $p_r$ and $r$ are defined in (\ref{rpr}), while $(M_0, \omega_0, {\mathcal{I}})$ is an arbitrary
Hamiltonian system on the symplectic  space  ($M_0$, $\omega_0$) parameterized by some set of
coordinates $\{u_1, u_2,\ldots\}$.

It is seen from the above consideration that  the whole information
about conformal system is encoded, in some way, in the the
lower-dimensional system $(M_0, \omega_0, {\mathcal{I}})$
In particular, it is obvious that if the system $(M_0, \omega_0,
{\mathcal{I}})$ is integrable in the sense of Liouville then the
corresponding conformal mechanics (\ref{generic}) is also integrable
in this sense \cite{burdik}. It is also obvious that the separation of variables
of the $(M_0, \omega_0, {\mathcal{I}})$ immediately implies the
separation of variables in the corresponding conformal mechanics.
Vice versa, having at hand the  exact solutions of the conformal
mechanics (\ref{generic}), we immediately get the explicit solutions
of the system $(M_0, \omega_0, {\mathcal{I}})$. Hence, the integrability
of the generic conformal mechanics leads to the integrability of
$(M_0, \omega_0, {\mathcal{I}})$. However, it is unclear, how to extract
the Liouville constants of motion of the angular part from ones of
the underlying conformal mechanics. More generally, it is unclear,
how the dynamical conformal symmetry impacts on the integrability
properties of the system.
In the next sections we present our preliminary observations on this subject.

An important particular case of the conformal system is the conformal mechanics
on the Euclidean space $\DR^N$,
\begin{equation}
\omega={d{\bf p} }\wedge {d{\bf r}},
\qquad
H= \frac{{\bf p}^2}{2}+V({\bf r}),
\quad
{\rm where}
\quad
{\bf r}\cdot\mathbf{\nabla} V({\bf r} )=-2V({\bf r}),
\label{h1}
\end{equation}
and ${\bf r}=(x_1,\ldots , x_N)$, ${\bf p}=(p_1,\ldots, p_N)$ are, respectively,  the Cartesian coordinates of $\DR^N$
and their conjugated momenta.
 In this  case  the generators $D$ and $K$ looks as follows
\begin{equation}
D={\bf p}\cdot{\bf r},\qquad  K=\frac{{\bf r}^2}{2}. \label{dk1}
\end{equation}
Extracting the radius $r=|{\bf r}|$, and introducing the conjugated momentum
 $p_r={\bf p}\cdot{\bf r}/r $,  we can represent the above
generators  as in \eqref{so2}.
In this terms the Casimir $\mathcal{I}$ is defined by the expression
\be
 \mathcal{I}\equiv \sum_{i<j}\frac{ l_{ij}^2}{2}+ U(\mathbf{r}),
 \qquad U(\mathbf{r})\equiv r^2V({\bf r}),
\ee
where $l_{ij}=p_ix_j-p_jx_i$ is the angular momentum.
The role of $M_0$ plays the cotangent bundle of $(N-1)$-dimensional sphere, $T^*S^{N-1}$.
In  the  spherical coordinate system with  $r,p_r$ taken as radial coordinates,
$\mathcal{I}$ depends only on the $N-1$ angular coordinates $\phi^\alpha $ and
their canonically conjugated momenta $\pi_\alpha$.
This particular Hamiltonian system can be presented in the form
\begin{equation}
 \left(T^*S^{N-1},\; \omega_0=d\pi_\alpha \wedge d\phi^\alpha,\;
 \mathcal{I}(u)=\tfrac{1}{2}{g^{\alpha\beta}(\phi)\pi_\alpha \pi_\beta} +U(\phi)\right ),
 \label{hs}
 \end{equation}
where $g^{\alpha\beta}(\phi)$ is the inverse metrics of the hypersphere, and $\alpha=1,\ldots, (N-1)$.
Its Hamiltonian,  being quadratic on momenta, describes a particle
moving on $S^{(N-1)}$  in the presence of potential $U(\phi)$.

 An interesting example of such a spherical system
is provided by  the rational $N-$particle Calogero model with the excluded  center-of-mass.
Its spherical part defines the $N(N-1)$-center Higgs oscillator on the
$(N-2)$-dimensional sphere with the oscillator centers located in the roots of the $A_{n-1}$ Lie algebra.
In the three-particle case, the oscillator centers are located at the vertexes of hexagon
while in the four-particle case, they correspond to the  vertexes of cuboctahedron \cite{cuboct}.

\setcounter{equation}{0}
\section{Integrals with certain conformal dimension}
The constant of motion of the conformal mechanics (\ref{so2}),
which is of the $n$-th order conformal dimension,
$\{D, I_n\}=-nI_n$, can be decomposed on $p_r, r$ as follows
\be
\label{In}
I_n(p_r,r, u )=\sum_{k=0}^n
\frac{p_r^{n-k}}{r^k}f_n^k(u),
\ee
where the coefficients $f_n^k(u)$ depend on the variables parameterizing
$M_0$, and  do not depend
on the radial coordinate $r$ and momentum $p_r$.
Here and in the following we will use the notation $\hat{X}f\equiv\poiss{X}{f}$
for the Poisson bracket action.

Note that for the conformal mechanics, any integral of motion can be
presented as a sum of integrals with certain conformal dimension.
This follows from the conformal invariance of the system \eqref{h1},
or the zero dimensionality of the coupling constant. Indeed, for any
integral presented in the form $I=\sum_iI_{n_i}$ we have
$\sum_i\hat{H}I_{n_i}=0$. Acting by $\hat{D}^k$ on both sides of
the last equation and using the commutation relation between
$\hat{D}$ and $\hat{H}$ inherited from \eqref{so12}, we arrive at
the another equation $\sum_i n_i^k\hat{H}I_{n_i}=0$ for any $k$.
This implies that the quantities $I_{n_i}$  conserve if their
conformal dimensions  differ. Therefore, the representation
\eqref{In} do not put any restriction on the constant of motion. 
Using the identity
$\hat{\mathcal{I}}=2(K\hat H+H \hat K)-D\hat
D$, we get
\be \label{hatI} \hat{\mathcal{I}}I_n=(2H\hat{K}+nD)I_n.
\ee
Inserting the decomposition \eqref{In},
we observe that the operator $\hat{\mathcal{I}}$ acts only on the $f_n^k(u )$,
while the operator $2H\hat{K}+nD$ acts only on the basic monomials
\be
\label{basis}
\frac{p_r^{n-k}}{r^k},\qquad k=0,1,\dots,n,
\ee
namely
\be
\label{defM}
(2H\hat{K}+nD)\frac{p_r^{n-k}}{r^k}
=k\frac{p_r^{n-(k-1)}}{r^{k-1}}-2(n-k)\mathcal{I}\frac{p_r^{n-(k+1)}}{r^{k+1}}.
\ee
It is easy to see that it is invariant on the space formed by the monomials
\eqref{basis} with $\mathcal{I}$-dependent coefficients. This enables us to obtain
the following recursion relation from \eqref{hatI}
\be
\label{recurrent}
\hat{\mathcal{I}} f^k_n=(k+1)f^{k+1}_n-2(n-k+1){\cal I}f^{k-1}_n,
\ee
where the ``boundary" conditions $f^{-1}_n=f^{n+1}_n=0$ are supposed.

This recursion relation can be  rewritten in the matrix form
\be
\label{matrix}
\hat{\mathcal{I}}\,\mathbf{f}_n=\mathbf{M}_n\mathbf{f}_n,
\ee
where  
\be
\label{Mn}
\mathbf{M}_n=
\left(
\begin{array}{cccccc}
0 & -1 & 0 & \ldots &0 & 0 \\
2n{\cal I}  & 0 & -2 & \ldots & 0 & 0 \\
0 & 2(n-1){\cal I}  & 0 &  \ldots & 0 & 0 \\
\dots&\dots&\dots& \dots& \dots& \dots \\
0 & 0 & 0 & \ldots &  0 & -n\\
0 & 0 & 0 & \ldots & 2{\cal I} &0
\end{array}
\right)
\qquad\textrm{and}\qquad
\mathbf{f}_n=
\left(
\begin{array}{c}
f_n^0 \\ f_n^1 \\ \dots \\ \dots \\ f_n^{n-1}\\f^n_n
\end{array}
\right).
\ee
The matrix $\mathbf{M}_n$ \eqref{Mn} is conjugated to the $x$-projection of
the spin-$\frac{n}{2}$ operator:
\be
 \mathbf{M}_n=-2\sqrt{-2\mathcal{I}}\ \mathbf{D}\mathbf{L}_x\mathbf{D}^{-1},
 \quad
 \text{where} \quad \mathbf{D}=\textrm{diag}(d_0,\dots,d_n),
 \quad
 d_k=\sqrt{\binom{n}{k}}(-2\mathcal{I})^{k/2},
\label{dec}\ee
and
${\bf L}_x$ is the matrix with the following nonzero elements
\be
(\mathbf{L}_x)_{M,M-1}=(\mathbf{L}_x)_{M-1,M}=\frac12\sqrt{(L+M)(L-M+1)},
\label{spin}\ee
i.e. is the  the $x$-projection of spin $L$ operator\cite{ll}.
So, the spectrum of $\mathbf{M}_n$ is proportional to the spectrum 
of $\mathbf{L}_x$.

The system of recursion relations  \eqref{matrix}
is overdefined.
The compatibility condition is equivalent to the requirement
\be
\label{Detfn}
\text{Det}(\mathbf{1}\cdot\hat{\mathcal{I}}-\mathbf{M}_n)\, f^i_n=0,
\ee
where $\mathbf{1}$ is the $(n+1)$-dimensional identity matrix.
Joining pairwise the monomials with the same absolute
value of the spin projection in order to avoid roots of $\mathcal{I}$,
we arrive at the following expression for the determinant:
\be
\label{det}
\text{Det}(\mathbf{1}\cdot\hat{\mathcal{I}}-\mathbf{M}_n)=
{\displaystyle
\begin{cases}
\prod_{k=0}^{n/2-1}(\hat{\mathcal{I}}^2 +2(n-2k)^2{\cal I})\hat{\mathcal{I}} & \text{for even $n$},
\\[5pt]
\prod_{k=0}^{(n-1)/2}(\hat{\mathcal{I}}^2 +2(n-2k)^2{\cal I}) & \text{for odd $n$}.
\end{cases}
}
\ee
It is obvious that any integral \eqref{In}  satisfies the similar equation
$\text{Det}(\mathbf{1}\cdot\hat{\mathcal{I}}-\mathbf{M}_n)\, I_n=0$.
Since the operator  (\ref{Mn}) clearly commutes with (\ref{det}), we conclude
that the set of equations (\ref{Detfn})
is overcompleted: the fulfilling of this equation for the
 $f^0_n(u)$,
\be
\text{Det}(\mathbf{1}\cdot\hat{\mathcal{I}}-\mathbf{M}_n)\, f^0_n=0,
\label{char}\ee
  leads the fulfilling of this equation for any $f^i_n(u)$.

Hence,
the coefficients $f_n^k(u)$ in the decomposition of $n$-th order integral
\eqref{In} are uniquely determined
by  $f_n^0(u)$, which describes  the asymptotical behavior of $I_n$
in the large radius limit $r\to\infty$.
Note that in the above derivations, the form of the angular part
${\cal I}$ is unessential. From the \eqref{det} we see that the
integral ${\cal I}$ is not in involution with any odd order integral
$I_{2k+1}$. Moreover,  for any (even) integral of motion the
condition $\hat{\mathcal{I}}^2I_{2k}=0$ can be fulfilled only if
$I_{2k}$ is in involution with $\mathcal{I}$. In other words, a
nontrivial action of the  operator $\hat{\mathcal{I}}$ on the
integrals of motion is not nilpotent.

Let us complete this section by the explicit solution of the
recursion relations \eqref{recurrent}. 
For this purpose let us present  \eqref{recurrent}
in the $2\times2$ matrix form:
\be
\label{Tmat}
\left(
\begin{array}{c}
f_n^{k} \\
f_n^{k-1}
\end{array}
\right)
=
\left(
\begin{array}{cc}
 \frac{1}{k}\hat{\mathcal{I}} & 2\mathcal{I}\frac{n-k+2}{k}
\\
1 & 0
\end{array}
\right)
\left(
\begin{array}{c}
f_n^{k-1} \\
f_n^{k-2}
\end{array}
\right).
\ee
Then one can present the solution of this system of equations
in terms of the matrix product:
\be
f^k_n=[\mathbf{T}_k\mathbf{T}_{k-1}\ldots\mathbf{T}_1]_{11} f_n^0,
\ee
where $\mathbf{T}_k$ is the $2\times2$ transfer matrix defined in \eqref{Tmat}.
From this equation, we see that $f^k_n$ is the following polynomial on
$\hat{\mathcal{I}}$ and $\mathcal{I}$:
\be
f_n^k=\sum_{i=0}^{[k/2]} c_{ki}(n)\,\mathcal{I}^i\hat{\mathcal{I}}^{k-2i}f_n^0,\quad {\rm where}\quad
c_{ki}(n)=\frac{1}{k!}\sum_{\substack{0<k_l<k \\ k_{l+1}-k_l>1}}
a_{k_1}(n)a_{k_2}(n)\dots a_{k_i}(n), \quad
a_k(n)=k(n-k+1).
\label{c}\ee
Thus, the above equations determine entirely the constant  of motion $I_n$ with conformal
dimension $n$ from its higher order coefficient $f_n^0(u)$
in $p_r$. The last function must obey the condition \eqref{Detfn}.
The equation \eqref{c} implies that the constant of motion $I_n$ vanishes if
it vanishes at the infinite boundary $r=\infty$.

\setcounter{equation}{0}
\section{Wojciechowski's
constants of motion}

By the use of operator $\hat{\mathcal{I}}$, we can get the
additional serie of constants of motion of the conformal mechanics,
associated with the known ones,
\be F_{k}\equiv
\hat{\mathcal{I}}I_k=\{{\cal I}, I_k\}:\quad \{F_{k}, {\cal H}\}=0.
\ee
These additional constants of motion precisely correspond to
the  hidden constants of motion of the Calogero model found by
Wojciechowski \cite{supint}.
For the self-consistency of the paper we present
this  construction in general terms.

Following \cite{supint}, we introduce the auxiliary  quantities
\be
\label{Jn}
J_n=\hat{K}I_n= -\sum_{k=0}^{n-1}(n-k)\frac{p_r^{n-k-1}}{r^{k-1}}f_k,
\ee
which evolution is  linear in time
\be
\dot{J_n}=\hat{H}J_n=\hat{H}\hat{K}I_n=\hat{D}I_n=-nI_n.
\ee
Taking into account this expression, it is easy to check that the following expression also defines the constant  of motion:
\be
\label{Knm}
L_{m,n}=mI_m J_n-nI_nJ_m,\qquad L_{n,m}=-L_{m, n}\; .
\ee
For the Calogero system, $L_{1,n}$ with $n=2,\dots,N$ give rise to additional $N-1$ integrals,
which form, together with Liouville integrals,  the functionally independent
set of  $2N-1$ constants of motion. This provides the Calogero model
with the maximal superintegrability property.
For the system with reduced center of mass, the first-order integral
$I_1$ describing the total momentum of the initial system, vanishes. The Liouville integrals
are obtained from the reduction of the Calogero integrals starting from $n=2$. The additional
$N-2$ integrals correspond to $L_{2,n}$ with $n\ge3$.

Turning back to the equation \eqref{hatI}, we see that
it also defines a constant of motion of $n$th conformal dimension,
which, of course, possesses the similar decomposition in powers of $p_r$ \eqref{In}.
Take now $H$ as a second order integral $I_2$, which implies $J_2=\{K,H\}=-D$.
Using the definition \eqref{Knm}, one can rewrite \eqref{hatI} as follows:
\be
\label{I-In'}
F_n\equiv\hat{\mathcal{I}}I_n=2I_2J_n-nI_nJ_2=L_{2,n}.
\ee
So, the Hamiltonian action of $\mathcal{I}$ on the  Liouville
constants of motion of the Calogero model
yields  the additional serie of the constants of motion found by Wojciechowski in \cite{supint}.

We proved that $\mathcal{I}$ is not in involution
with any Liouville integrals except the Hamiltonian itself because
the integrals $L_{2,n}$, $n\ne2$ do not vanish and form together with Liouville
ones a complete set of functionally independent constants of
motion \cite{supint}.
Using \eqref{Knm}, one can  derive a more general relation
 between $\mathcal{I}$ and other integrals of motion:
\be
\label{I-In}
mI_m\poiss{\mathcal{I}}{I_n}-nI_n\poiss{\mathcal{I}}{I_m}=2HL_{m,n}.
\ee
Note that, for any two constant of motion $I_n$ and $I_m$, the quantity
$I_{n.m}\equiv\poiss{I_n}{I_m}$ also conserves and has the conformal
dimension $n+m$ provided that it does not vanish since
\be
\hat{D}{I_{n.m}}=\poiss{\hat{D}I_n}{I_m}+\poiss{I_n}{\hat{D}I_m}=
-(n+m)I_{n.m}.
\ee
Therefore, the decomposition \eqref{In}
with the replacement $n\to n+m$ is valid for $I_{n. m}$ too.
The higher order coefficient in $p_r$ is simply given
by $f_{n+m}^0=\poiss{f_n^0}{f_m^0}$.
The product of two integrals $I'_{n+m}=I_nI_m$ has the same property.
It conserves, has the same conformal dimension, and
its higher order coefficient coincides with the product
$f_{n+m}'^0=f_n^0 f_m^0$.

Thus, if some set of constants of motion $\{I_n\}$  form closed algebra
with respect to the Poisson brackets, then their leading asymptotes
$\{f_n^0\}$ will form the same algebra. The contraction map
$I_n\to f_n^0=\lim_{r\to\infty}I_n$ defines a one-to-one correspondence 
conserving the Poisson structure.
In particular, if the constants of motion are
in involution, i.e.
\be\{I_n,I_m\}=0 ,
\ee
then their  limits
at infinite radius are subjected to the
similar relation, i.e. $\{f_n^0,f_m^0\}=0$, and vice versa.
The converse assertion can be also checked by direct
calculation using the recursion relations \eqref{recurrent}.
Hence, the analysis of the constants of motion of the conformal mechanical
system can be reduced to the study of the equation (\ref{char}).\\

\noindent
{\bf Example: \ }
As simplest example, 
let us consider the three-particle Calogero model with excluded center of mass.
The angular part of its Hamiltonian  takes rather simple form \cite{jacobi}
\be
{\cal I}=\frac{p_{\varphi}^2}{2}+\frac{9g}{2\cos^2 3\varphi},
\ee
and describes  particle motion on a circle interacting with three equidistant
Higgs oscillator centers.
The rest constants of motion in  polar coordinates look as follows
\cite{cuboct}:
\be
I_3=
   \left(p_r^2 - \frac{6 {\cal I}}{r^2}\right)  p_r \sin 3 \varphi +
   \left(3p_r^2 - \frac{2 {\cal I}}{r^2}\right)  \frac{p_\varphi \cos 3\varphi}{r},
\quad F_3=
     \left(p_r^2 - \frac{6{\cal I}}{r^2} \right) 3p_r p_\varphi \cos 3\varphi
     -  \left(3p_r^2 - \frac{2{\cal I}}{r^2} \right)\frac{6{\cal I}\sin 3 \varphi}{r}.
\ee
These four constants of motion $\mathcal{I},H,I_3,F_3$ are functionally
dependent with the following algebraic relation:
$
 F_3^2 + 2{\cal I} I^2_3= 8 H^3(2{\cal I}-9g)$.
They  generate the algebra with the following nonvanishing brackets: \be
\label{adI} \{{\cal I},I_3\}=3 F_{3}, \qquad \{{\cal I},F_3\}=-6
{\cal I} I_3,\qquad \{{F_3,I_3}\}=3(8  H^3- {I}^2_3). \ee
As was proved above for generic conformal mechanics
(see \eqref{Detfn}, \eqref{det}),
the operator
\be
\label{2-18}
{\bf M}_3=(\hat{\mathcal{I}}^2+2{\cal I})(\hat{\mathcal{I}}^2+18{\cal I})
\ee
must annihilate any third order constant of motion.
We see from \eqref{adI} that just its factor monomial
$\hat{\mathcal{I}}^2+18{\cal I}$ vanishes on
both third order  integrals $I_3$ and $F_3$, i.e., in fact,
a stronger condition is fulfilled in this case.
Since  $(T^* S^1, \omega_0, {\cal I})$ defines one-dimensional Hamiltonian system, it has no any
nontrivial constants of motion.

\setcounter{equation}{0}
\section{Constants of motion for $(M_0, \omega_0, \mathcal{I})$}
We demonstrated, in the previous  section, that the information on the conformal mechanics is encoded
 in its angular part $(M_0, \omega_0, \mathcal{I})$, which could be considered, by itself,
as some Hamiltonian system.
It is obvious that the integrability of the initial conformal mechanics leads to the integrability of
the ``angular system" $(M_0, \omega_0, \mathcal{I})$, and vice versa.
It is also evident that the constants of motion of the angular system 
are constants of motion for the conformal mechanics. 
Yet, the inverse statement is not true, though
clearly, there should be some relation between the constants of motion of 
the conformal mechanics and those of its angular part.

So how to construct the constants of motion of $(M_0, \omega_0, \mathcal{I})$ 
from the ones of the initial conformal mechanics?
For any integral $I_n$ of the Hamiltonian \eqref{generic}
with even conformal dimension, an integral of $\mathcal{I}$ can be constructed
from its coefficients in the decomposition \eqref{In} easily.
Indeed, the matrix $\mathbf{M}_n$ defined in \eqref{Mn} is singular 
for even $n$, as follows from \eqref{dec} and \eqref{spin}.
Denote by $\mathbf{g}_n$ the singular vector of the transposed matrix:
$\mathbf{M}_n^\tau\mathbf{g}_n=\mathbf{g}_n^\tau\mathbf{M}_n=0$.
Then, from \eqref{matrix} we obtain
\be
\hat{\mathcal{I}}(\mathbf{g}_n^\tau\mathbf{f}_n)
=
\mathbf{g}_n^\tau\hat{\mathcal{I}}(\mathbf{f}_n)
=
\mathbf{g}_n^\tau\mathbf{M}_n\mathbf{f}_n=0,
\ee
which means that $G_n=\mathbf{g}_n^\tau\mathbf{f}_n$ is an integral of $(M_0, \omega_0, \mathcal{I})$
(if it does not vanish).
Calculating the coordinates of the singular vector, we obtain the following explicit
form of the integral:
\be
G_n=\sum_{i=0}^{n/2}(2i-1)!!(n-2i-1)!!(2\mathcal{I})^{-i} f_n^{2i}=
\sum_{i=0}^{n/2}\sum_{j=0}^{i}\alpha_{ij}(n)\mathcal{I}^{j-i} \hat{\mathcal{I}}^{2(i-j)}f_n^0
\label{si}\ee
where
\be
\alpha_{ij}(n)=2^{-i}(2i-1)!!(n-2i-1)!!\,c_{2i, j}(n),\qquad (-1)!!=1
\ee and  $c_{2i,j}(n)$ is defined by \eqref{c}.
As was proved above, the operator \eqref{det} must annihilate $f_n^0$. It is a $(n+1)$th order
polynomial on $\hat{\mathcal{I}}$ while the expression of $G_n$ contains $n$ and less orders of
$\hat{\mathcal{I}}$. Therefore, the mentioned condition does not put any constraint on the
terms, and there is a hope that this constant of motion
does not vanish for concrete models.

The relation of the constants of motion of the
$(M_0, \omega_0, \mathcal{I})$, which are associated with the ones of initial conformal mechanics,
with odd conformal dimensions is more complicated.

\setcounter{equation}{0}
\section{${\cal N}{=}4$ superconformal mechanics}
In the previous sections we demonstrated that the analysis of conformal mechanics can be reduced
to the study of its ``angular part" given by the lower-dimensional Hamiltonian system
$(M_0, \omega_0, \mathcal{I})$. Vise versa, such a system can be easily lifted
to some conformal mechanics.
In this section we will show that a similar correspondence exists between 
${\cal N}{=}4$ superconformal mechanics with symmetry algebra $D(1,2|\alpha)$
its and ${\cal N}{=}4$ supersymmetric angular part. Namely, any  ${\cal N}{=}4$ supersymmetric extention
of the ``angular" system
$(M_0, \omega_0, \mathcal{I})$ can be lifted to the $
D(1,2|\alpha)$ \cite{paul} superconformally invariant one  by coupling with super dilaton.

To describe this procedure let us, at first, consider the realization of the $D(1,2|\alpha)$ superalgebra on the $(2|4)$-dimensional
 phase superspace parameterized by the coordinates $r,p_r, \eta^{a\alpha}$, and equipped with the canonical Poisson brackets
 \be
 \{p_r, r\}=1,\quad \{\eta^{a\alpha},\eta^{b\beta}\}=2\imath\varepsilon^{ab}\varepsilon^{\alpha\beta},
 \ee
where all indices run from 1 to 2.
We define, on this phase space, the following generators

\be
D=p_r r, \quad K=\frac{r^2}{2},\quad {\cal H}_0=\frac{p_r^2}{2}+\frac{(1+2\alpha )\eta^{a\beta}\eta_{b\beta}\eta^{b\alpha}\eta_{a\alpha}}{12r^2},
\quad V^{ab}_0=\eta^{a\alpha}\eta^{b}_{\alpha},\quad  W^{\alpha\beta}_0=\eta^{a\alpha}\eta_{a}^{\beta}.
\ee
\be
S^{a\alpha}=r\eta^{a\alpha},\quad Q_0^{a\alpha}=p_r\eta^{a\alpha}-\frac{(1+2\alpha )\eta^{a\beta}\eta_{b\beta}\eta^{b\alpha}}{3r},
\ee
These generators form the superalgebra $D(1,2;\alpha )$, given by the following non-zero relations
\bea
&\{ {\cal H}_0 , D\}= 2 {\cal H}_0,  \quad  \{ K, D\}=-2 K, \quad\{  {\cal H}_0 , K\}=D.&\nonumber\\
&\{{V}^{ab}_0, { V}^{cd}_0\}=2\imath\left(\varepsilon^{ac}{ V}^{bd}_0+
\varepsilon^{bd}{ V}^{ac}_0
\right),\;\;
\{{ W}^{\alpha\beta}_0, {W}^{\gamma\delta}_0\}=2\imath\left(\varepsilon^{\alpha\gamma}{ W}^{\beta\delta}_0
+\varepsilon^{\beta\delta}{ W}^{\alpha\gamma}_0\right),
&\nonumber\\
&\{Q^{a\alpha}_0, Q^{b\beta}_0\}=2\imath\varepsilon^{ab}\varepsilon^{\alpha\beta}{\cal H}_0,\quad
\{S^{a\alpha}, S^{b\beta}\}=2\imath\varepsilon^{ab}\varepsilon^{\alpha\beta}K,
&\nonumber\\
&\{Q^{a\alpha}_0, S^{b\beta}\}=2\imath\varepsilon^{ab}\varepsilon^{\alpha\beta}D
+\alpha \varepsilon^{ab}W^{\alpha\beta}_0 -(1+\alpha )\varepsilon^{\alpha\beta}V^{ab}_0 ,
&\nonumber\\
&\{S^{a\alpha},{\cal H}_0\}=-Q^{a\alpha},\quad    \{D, Q^{a\alpha}_0\}=-Q^{a\alpha}_0,\quad
\{Q^{a\alpha_0},K\}= S^{a\alpha},\quad \{D, S^{a\alpha}\}= S^{a\alpha}&\nonumber\\
&\{{V}^{ab}_0,
Q^{c\alpha}_0\}=\imath\left(\varepsilon^{ac}Q^{b\alpha }_0+
\varepsilon^{bc}Q^{a\alpha}_0 \right),\quad \{{ W}^{\alpha\beta}_0,
Q^{a\gamma}_0\}=\imath\left(\varepsilon^{\alpha\gamma}Q^{a\beta}_0
+\varepsilon^{\beta\gamma}Q^{a\alpha}_0\right),&\nonumber \\
&\{{ V}^{ab}_0, S^{c\alpha}\}=\imath\left(\varepsilon^{ac}S^{b\alpha }+ 
\varepsilon^{bc}S^{a\alpha} \right),\quad
\{{ W}^{\alpha\beta}_0, S^{a\gamma}\}=\imath\left(\varepsilon^{\alpha\gamma}S^{a\beta}
+\varepsilon^{\beta\gamma}S^{a\alpha}\right).&\label{100}
\eea
Notice that for the $\alpha=0,-1$ we get the $su(1,1|2)$ superalgebra 
while for $\alpha=-1/2$ emerges the $osp(4|2)$ one.

Now, we extend this system to the conformal mechanics with angular part
 ${\cal I}_{SUSY}={\cal I}+\ldots$, which  is the ${\cal N}{=}4 $ supersymmetric ``angular Hamiltonian"
${\cal I}$.
\bea
&\{\Theta^{a\alpha},\Theta^{b\beta}\}=2\imath\varepsilon^{ab}\varepsilon^{\alpha\beta} {\cal I}_{SUSY},
\qquad \{\Theta^{a\alpha},{\cal I}_{SUSY}\}=0,&\nonumber\\
& \{V^{ab}_1, \Theta^{c\alpha}\}=-\imath \left(\varepsilon^{ac}\Theta^{b\alpha}+
 \varepsilon^{bc}\Theta^{a\alpha}\right),\quad \{W^{\alpha\beta}_1, \Theta^{a\gamma}\}=
-\imath \left(\varepsilon^{\alpha\gamma}\Theta^{a\beta}+
\varepsilon^{\beta\gamma}\Theta^{a\alpha}\right),&\\
&\{V^{ab}_1, {\cal I}_{SUSY}\}=\{W^{\alpha\beta}_1,
 {\cal I}_{SUSY}\}=0.&\nonumber
\eea
where $V^{ab}_1$, $W^{\alpha\beta}_1$ form $su(2)\times su (2)$ algebra
\be
\{V^{ab}_1, V^{cd}_1\}=2\imath\left(\varepsilon^{ac}V^{bd}_1+\varepsilon^{bd}V^{ac}_1\right),\quad
\{W^{\alpha\beta}_1, W^{\gamma\delta}_1\}=2\imath\left(\varepsilon^{\alpha\gamma}W^{\beta\delta}_1
+\varepsilon^{\beta\delta}W^{\alpha\gamma}_1\right), \quad \{V^{ab}_1,W^{\alpha\beta}_1\}=0.
\ee
Surely, $V^{ab}_1$, $W^{\alpha\beta}_1$ are precisely the $R$-symmetry generators of ${\cal N}{=}4$ supersymmetry.
We give some particular realization of these generators via supercharges $\Theta^{a\alpha}$ and Hamiltonian ${\cal I}_{SUSY}$:
\be
V^{ab}_1=\frac{\Theta^{a\alpha}\Theta^{b}_{\alpha}+ \Theta^{b\alpha}\Theta^{a}_\alpha }{2{\cal I}_{SUSY}},\quad
W^{\alpha\beta}_1=\frac{\Theta^{a\alpha}\Theta_{a}^\beta+ \Theta^{a\beta}\Theta_{a}^\alpha }{2{\cal I}_{SUSY}}.
\ee

Let us assume that ``radial'' variables $p_r, r, \eta^{a\alpha}$ commute  
with all elements of the above ${\cal N}{=}4$ superalgebra (\ref{100}).
Then, we define  the following  odd generators
\be
 S^{a\alpha}=r\eta^{a\alpha},\qquad Q^{a\alpha}=Q^{a\alpha}_0+\frac{\Theta^{a\alpha}}{r} -
\imath \alpha \frac{\eta^{a}_{\beta}W^{\beta\alpha}}{r}
+\imath (1+\alpha )\frac{\eta^{\alpha}_{b}V^{ba}}{r},
\label{sch}\ee
and the even ones
\be
{\cal H}= {\cal H}_0 +\frac{{\cal I}_{SUSY}}{r^2}-\frac{\imath\Theta^{a\alpha}\eta_{a\alpha}}{r^2}
+\frac{(1+\alpha )V^{ab}_1V_{(0)ab}}{2r^2}-
\frac{\alpha W^{\alpha\beta}_1W_{(0)\alpha\beta}}{4r^2}
+\frac{(1+2\alpha )}{12r^2}
\left( V^{ab}_1V_{(1)ab}+W^{\alpha\beta}_1W_{(1)\alpha\beta}\right),\nonumber\ee
\be
K=\frac{r^2}{2},\quad D= p_r r, \qquad { V}^{ab}=V^{ab}_1+V^{ab}_0,
\quad {W}^{\alpha\beta}=W^{\alpha\beta}_1+W^{\alpha\beta}_0.
\label{suh}\ee

It is easy to check that these new generators obey the relations \eqref{100},
where the following replacement is made
\be
Q^{a\alpha}_0\to Q^{a\alpha},\quad  {\cal H}_0\to{\cal H},\quad V^{ab}_0\to { V}^{ab},
\quad W^{\alpha\beta}_0\to {W}^{\alpha\beta}
\ee
Hence, we presented the explicit construction which extends the 
arbitrary ${\cal N}{=}4$ supersymmetric Hamiltonian system 
$(M_0, \omega_0, \mathcal{I})$, to the  superconformal mechanics with
$D(1,2;\alpha)$ superconformal algebra.

 One should note that the proposed construction coupled ``angular'' part with ${\cal N}{=}4$ supermultiplet  $(1,4,3)$ \cite{Igreat}.
 The unique physical bosonic component of this supermultiplet   is just a dilaton field. Clearly enough, one may
 construct other variants of superconformal systems by a proper coupling of the ${\cal N}{=}4$ supersymmetric ``angular'' part ${\cal I}_{SUSY}$
with any supermultiplet containing the dilaton field, 
i.e.~with $(2,4,2)$, $(3,4,1)$ or $(4,4,0)$ ones. The explicit construction
 of the supercharges and Hamiltonian for such systems will be present elsewhere.

\section{Conclusion}
In present paper we split the generic conformal mechanics into a ``radial" 
and an ``angular" part, where the latter is a Hamiltonian system on the orbit
of $so(1,2)$, with the Hamiltonian given by the Casimir function of $so(1,2)$.
We reduced the analysis of the constants of motion of the conformal mechanics 
to the study of some differential equations on the orbit. Moreover, we found
that the separation of the system into radial and angular parts yields a 
transparent explanation of Wojciechowski's construction of the hidden constants
of motion of the rational Calogero model.
More precisely, the hidden constants of motion are simply the Poisson brackets 
of the Casimir function of $so(1,2)$ with the Liouville constant of motion of 
the full Calogero model.

This construction is valid for any conformal mechanics. 
Consequently, any integrable conformal mechanical system is superintegrable.
We also presented an explicit construction of the ``angular'' constants 
of motion from those of the full conformal mechanics.
Finally, we demonstrated how to lift an arbitrary ${\cal N}{=}4$ supersymmetric 
``angular" Hamiltonian system to a superconformal mechanics 
model with $D(1,2;\alpha )$ symmetry algebra.  

Various interesting issues remain to be tackled.
Simplest among them, and probably most important, is the quantum counterpart 
of our construction, including the spectrum and wavefunctions of the 
``angular part" of the rational Calogero model from the data of the 
initial system.

\acknowledgments{
T.~H. and A.~N. are grateful to Vahagn Yeghikyan for the useful comments. 
A.~N. acknowledges the hospitality in Dubna, 
where the part of this work has been done.
The work was supported in part by  RFBR 08-02-90490-Ukr, 06-16-1684(S.~K.), 
DFG 436 Rus 113/669/03 (S.~K., O.~L),
ANSEF PS1730 (A.~N.) and NFSAT-CRDF UC-06/07 (T.~H, A.~N) grants.}

\end{document}